\documentclass[aps,pra,twocolumn,showpacs,groupedaddress]{revtex4}
\usepackage{graphicx}

\newcommand{\ket}[1]{|#1\rangle}
\newcommand{\bra}[1]{\langle#1|}

\begin{document}

\title{Non-Markovian Effects on the  Geometric Phase}
\author{X. L. Huang and X. X. Yi }
\affiliation{School of physics and optoelectronic technology,\\
Dalian University of Technology, Dalian 116024 China }

\date{\today}

\begin{abstract}
The geometric phases of a two-level atom interacting with
non-Markovian environments are calculated and the non-Markovian
effects on the geometric phases are discussed in this paper. Three
kinds of methods that describe the non-Markovian process, projection
superoperator technique, memory kernel master equation and
post-Markovian master equation, are used in the discussions. The
results show that when the dissipation rate is large, the
non-Markovian effects change the geometric phase more strikingly
than the small one.

\end{abstract}

 \maketitle

\section{Introduction}
The so-called Berry phase or geometric phase was first pointed out
by Berry\cite{Berry} when he studied a pure quantum system that
undergoes an adiabatic cyclic evolution. Simon gave an
interpretation of this phase in the language of differential
geometry and fibre bundles\cite{Simon1983}. Since then, large
numbers of work have been done to generalize the conception of Berry
phase. For example, Aharonov and Anandan extended the Berry phase to
the case of non-adiabatic evolution\cite{AAnonadibatic},   Samuel
and Bhandari \cite{SBnoncyclic} generalized the Berry phase in
cyclic evolution to the case of non-cyclic evolution.

Another direction of the generalization is to study the geometric
phase for mixed states and nonunitary evolution. This work was first
done by Uhlmann\cite{Uhlmann} via the mathematical context of
purification. Sj\"{o}qvist {\it et at.} gave an alternate definition
of geometric phase for nondegenerate density operators undergoing
unitary evolution based on quantum
interferometry\cite{mixedsjoqvist}. This study has been extended to
the case of degenerate density operators by Singh {\it et
al.}\cite{mixedtong}. Geometric phase in nonunitary evolution have
been addressed in Ref.\cite{nonunitary1,nonunitary2}. Recently a
kinematic approach to the geometric phase for mixed quantal states
in nonunitary evolution is proposed by Tong {\it et
al.}\cite{TongDMGP}. The conception of off-diagonal geometric phase
for pure states was proposed in Ref.\cite{offdiagonalpure} and has
been extended to mixed states in Ref.\cite{offdiagonalmixed}. The
relation of geometric phase between entangled system and its
subsystem has been studied in Ref.\cite{entangledsub}.

The intrinsic nature of geometric phase provides us an impelling
tool for fault tolerance quantum computation\cite{faulttolerance}.
However, the unavoided  interaction between system and environment
would destroy the coherence of the quantum system and hence  to
limit the implementation of  quantum computation. So the study of
environment effects on geometric phase is highly required. Geometric
phases in dephasing system and spin environment were studied in
Ref.\cite{GPspin}, which used the exact solution of the system
dynamics and then it is fully non-Markovian. However the exactly
solvable models for open system are only few, most problems are
treated with certain approximations. For instance, the finite
temperature effect of environment on the mixed state geometric
phases\cite{GPtemperature} and the effects of a squeezed vacuum
reservoir on geometric phases\cite{GPsqueezedvac} were studied
within the Markovian approximation. It is well known that when the
dissipation is large or reservoir is finite and the initial state of
the composite system(system and environment) is entangled, the
process exhibits strong non-Markovian effects. In this paper we
shall study the non-Markovian effects on the geometric phase of a
two-level atom interacting with a non-Markovian environment. Three
kinds of methods which describe the non-Markovian process,
projection superoperator technique, memory kernel master equation
and post-Markovian master equation, are considered. We calculate the
geometric phase for the system governed by these equations and
discuss the results obtained with different dissipation rates and
different memory effects.

Throughout this paper, the geometric phases are calculated according
to the formula of Tong\cite{TongDMGP},
\begin{widetext}
\begin{equation}
\Phi_{GP}(\tau)=\texttt{Arg}\left(\sum_{i=1}^N\sqrt{\lambda_i(0)\lambda_i(\tau)}\langle\varphi_i(0)|
\varphi_i(\tau)\rangle
\exp\left\{-\int_0^{\tau}\langle\varphi_i(t)|\dot{\varphi}_i(t)\rangle
dt\right\}\right),\label{TongGP}
\end{equation}
\end{widetext}
where $\lambda_i(t)$ are instantaneous eigenvalues of the density
matrix $\rho(t)$ and $|\varphi_i(t)\rangle$ are corresponding
eigenstates. The dot on $\varphi_i(t)$ denotes the derivative with
respect to time $t$, $\texttt{Arg}$ expresses the argument of a
complex number.

The structure of this paper is organized as follows. In Sec.II we
discuss a two-level atom interacting with a two-band environment.
The evolution equation is obtained via the projection superoperator
methods. In Sec.III, we calculate the geometric phase in an open
system governed by the memory kernel master equation and in Sec.IV
governed by the post-Markovian master equation. Conclusions and
discussions are presented in Sec.V

\section{non-Markovian effects with generalized Lindblad master equation}
In this section, we will calculate the geometric phase of a
two-level atom(system) coupled to a two-band environment in both
Markovian and non-Markovian cases. The evolutions of the reduced
density matrix for the atom in both cases are obtained through
correlated projection superoperators
method\cite{projection1,projection2,projection3,projection4,projectionmodel},
different chosen projection operators would differentiate the
Markovian and non-Mardovian cases. To calculate the geometric phase
we solve the time evolution of the system in the interaction picture
then back to the Schr\"{o}dinger picture according to the free
Hamiltonian of the system $H_S=\frac12\omega\sigma_z$. In the
following discussion, the initial state is chosen as,
\begin{equation}
|\psi(0)\rangle=\cos\frac{\theta}2\ket{1}+\sin\frac{\theta}2e^{i\phi}\ket{0}.\label{initialstate}
\end{equation}
Where $\ket{1}$ and $\ket{0}$ are the excited state and ground state
of the atom respectively. This state corresponds to a state vector
in Bloch sphere with polar angle $\theta$ and azimuthal angle
$\phi$.

\subsection{Model}

\begin{figure}
\includegraphics*[bb=1 1 184 166, width=6cm]{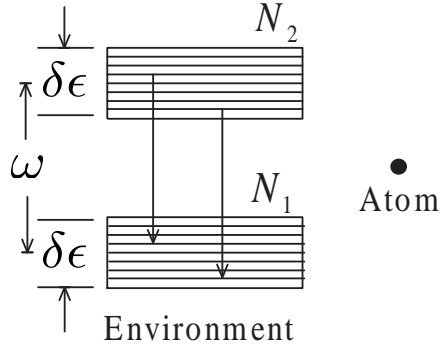}\caption{A two-level atom coupled to an
environment consisting of two energy bands with a finite number of
levels.} \label{setup}
\end{figure}

Consider a  two-level atom coupled to a two-band environment $E$.
The environment consists of a large number of energy levels which
are arranged in two energy bands with the same width(see
Fig.\ref{setup}). The levels of each band are equidistant. The lower
energy band contains $N_1$ levels while the upper band $N_2$ levels.
The transition of the atom is in resonance with distance $\omega$
between the bands (we set $\hbar=1$). The total Hamiltonian for such
a system is \cite{projectionmodel} $H=H_S+H_E+V$, with
\begin{eqnarray}
H_S=\frac12\omega\sigma_z,\nonumber
\end{eqnarray}
\begin{eqnarray}
H_E=\sum_{n_1}\frac{\delta\epsilon}{N_1}n_1|n_1\rangle\langle
n_1|+\sum_{n_2}(\omega+\frac{\delta\epsilon}{N_2}n_2)|n_2\rangle\langle
n_2|,\nonumber
\end{eqnarray}
\begin{eqnarray}
V=\lambda\sum_{n_1n_2}c(n_1,n_2)\sigma^+|n_1\rangle\langle
n_2|+\texttt{H.c.}
\end{eqnarray}
where the index $n_1$ denotes the levels of lower energy band and
$n_2$ the levels of upper band, $\sigma_z$ and $\sigma^+$ are Pauli
operators, $\lambda$ is the overall strength of the interaction,
$c(n_1,n_2)$ are coupling constants which are independent to each
other and satisfying,
\begin{eqnarray}
\langle c(n_1,n_2)\rangle&=&0,\nonumber\\
\langle c(n_1,n_2)c(n_1',n_2')\rangle&=&0,\nonumber\\
\langle
c(n_1,n_2)c^*(n_1',n_2')\rangle&=&\delta_{n_1,n_1'}\delta_{n_2,n_2'}.
\end{eqnarray}

Setting $H_0=H_S+H_E$, we transform our discussion to the
interaction picture that,
\begin{eqnarray}
V(t)=\sigma^+B(t)+\sigma^-B^{\dag}(t),
\end{eqnarray}
where
\begin{eqnarray}
B(t)=\lambda\sum_{n_1,n_2}c(n_1,n_2)e^{-i\omega(n_1,n_2)t}|n_1\rangle\langle
n_2|,
\end{eqnarray}
and
\begin{eqnarray}
\omega(n_1,n_2)=\delta\epsilon\left(\frac{n_2}{N_2}-\frac{n_1}{N_1}\right).
\end{eqnarray}

\subsection{Markovian case}

When the standard projection is used, the evolution of the reduce
system must be Markovian\cite{projection1,projection2,projection3}.
If we project the composite system's state into the form $\mathcal
{P}\rho=(\texttt{Tr}_E\rho)\otimes\rho_E=\rho_S\otimes\frac1{N_1}\Pi_1$,
where $\mathcal {P}$ is the projection superoperator and
$\Pi_1=\sum_{n_1}\ket{n_1}\bra{n_1}$, the second-order Markovian
master equation in the interaction picture is
obtained\cite{projectionmodel}:
\begin{equation}
\frac d{dt}\rho_S(t)=
\gamma_2\left(\sigma^-\rho_S(t)\sigma^+-\frac12\{\sigma^+\sigma^-,\rho_S(t)\}\right),\label{projectMarkovianME}
\end{equation}
where
\begin{eqnarray}
\gamma_2=\frac{2\pi\lambda^2N_2}{\delta\epsilon}.\nonumber
\end{eqnarray}

This equation is the same as that describes a two-level atom
spontaneously decay in vacuum under the Markovian approximation. The
time dependent of the reduce system in the Schr\"{o}dinger picture
may be obtained easily by a simple calculation. With the initial
state Eq.(\ref{initialstate}), the final state follows,

\begin{eqnarray}
\rho(t)=\left(\matrix{ \cos^2(\frac{\theta}{2})e^{-\gamma_2 t}   &
\frac12\sin\theta e^{-i(\phi+\omega t)}e^{-\frac12\gamma_2 t}   \cr
               \frac12\sin\theta e^{i(\phi+\omega
t)}e^{-\frac12\gamma_2 t} & 1-\cos^2(\frac{\theta}2)e^{-\gamma_2 t}
\cr
             }\right).
\end{eqnarray}

The two eigenvalues of $\rho(t)$ are
\begin{eqnarray}
\lambda_{\pm}=\frac12(1\pm\eta),\label{eigenvalueMarkovianP}
\end{eqnarray}
where
\begin{eqnarray}
\eta=\sqrt{\left(1-2\cos^2\frac{\theta}2e^{-\gamma_2
t}\right)^2+\sin^2\theta e^{-\gamma_2 t}}.
\end{eqnarray}
 It is obviously that the eigenvalue $\lambda_-=0$ at $t=0$. From
 the formation of Eq.(\ref{TongGP}), we can see that the eigenvalue
 $\lambda_-$ and its corresponding eigenstate $\ket{-}$ contribute
 zero to the geometric phase. This simplifies  the calculation.
 The eigenstate correspond to $\lambda_+$ can be written as:
 \begin{eqnarray}
 \ket{+}=\sin\frac{\theta_t}2\ket{1}+\cos\frac{\theta_t}2e^{i(\phi+\omega
 t)}\ket{0},\label{eigenstateMarkovianP}
 \end{eqnarray}
 where
 \begin{eqnarray}
\tan\frac{\theta_t}2=\frac{\sin\theta e^{-\frac12\gamma_2
t}}{1+\eta-2\cos^2\frac{\theta}2e^{-\gamma_2
t}}.\label{tanMarkovianP}
 \end{eqnarray}
Obviously, for $t=0$, $\tan\frac{\theta_t}2=\cot\frac{\theta}2$, as
expected. Now we substitute Eqs.(\ref{eigenvalueMarkovianP}) and
(\ref{eigenstateMarkovianP}) into Eq.(\ref{TongGP}) to obtain the
geometric phase at time $t$,
\begin{widetext}
\begin{eqnarray}
\Phi_{GP}=\texttt{Arg}\left[\left(\cos\frac{\theta}2\sin\frac{\theta_t}2
+\sin\frac{\theta}2\cos\frac{\theta_t}2e^{i\omega
t}\right)\exp\left(-i\omega\int_0^t\cos^2\frac{\theta_t}2dt\right)\right].
\end{eqnarray}
\end{widetext}

We consider the geometric phase acquired after a quasi-periods
$T=\frac{2\pi}{\omega}$. In this case, we can rewrite the geometric
phase as
\begin{eqnarray}
\Phi_{GP}=-\omega\int_0^T\cos^2\frac{\theta_t}2dt,\label{GP}
\end{eqnarray}
where $\cos^2\frac{\theta_t}2$ can be obtained according to
Eq.(\ref{tanMarkovianP})
\begin{eqnarray}
\cos^2\frac{\theta_t}2=\frac{(1+\eta-2\cos^2\frac{\theta}2e^{-\gamma_2
t})^2}{(1+\eta-2\cos^2\frac{\theta}2e^{-\gamma_2 t})^2+\sin^2\theta
e^{-\gamma_2 t}}.
\end{eqnarray}

\subsection{Non-Markovian case}

If we choose a correlated projection on the total system
\begin{eqnarray*}
\mathcal
{P}\rho&=&\texttt{Tr}_E(\Pi_1\rho)\otimes\frac1{N_1}\Pi_1+\texttt{Tr}_E(\Pi_2\rho)\otimes\frac1{N_2}\Pi_2\\
&=&\rho_S^{(1)}\otimes\frac1{N_1}\Pi_1+\rho_S^{(2)}\otimes\frac1{N_2}\Pi_2,
\end{eqnarray*}
where $\Pi_2=\sum_{n_2}\ket{n_2}\bra{n_2}$, the evolution of the
reduced system is said to be
non-Markovian\cite{projection1,projection2,projection3}. At this
stage the master equation in the interaction picture
is\cite{projectionmodel}
\begin{eqnarray}
\frac d{dt}\rho_S^{(1)}(t)=
\gamma_1\sigma^+\rho_S^{(2)}(t)\sigma^--\frac{\gamma_2}2\{\sigma^+\sigma^-,\rho_S^{(1)}(t)\},\nonumber \\
\frac d{dt}\rho_S^{(2)}(t)=
\gamma_2\sigma^-\rho_S^{(1)}(t)\sigma^+-\frac{\gamma_1}2\{\sigma^-\sigma^+,\rho_S^{(2)}(t)\},
\end{eqnarray}
where $\gamma_1$ has the similar definition to $\gamma_2$ and the
state of atom is given by $\rho_S=\rho_S^{(1)}+\rho_S^{(2)}$. This
equation is of a generalized Lindblad form\cite{projection2} and
gives an excellent approximation of the reduced system's
dynamics\cite{projectionmodel}.

We assume that initially only the lower band is populated. This
means $\rho^{(2)}_S(0)=0$. We also set $\gamma_1=\gamma_2=\gamma$.
The evolution of the atom with the initial condition
(\ref{initialstate}) in the Schr\"{o}dinger picture is
\begin{eqnarray}
\rho(t)=\left(\matrix{\frac12(1+e^{-2\gamma t})\cos^2\frac{\theta}2
& \frac12\sin\theta e^{-i(\phi+\omega t)}e^{-\frac12\gamma t} \cr
 \frac12\sin\theta e^{i(\phi+\omega t)}e^{-\frac12\gamma t} & 1-\frac12(1+e^{-2\gamma
 t})\cos^2\frac{\theta}2\cr}\right).
\end{eqnarray}

Following the same procedure mentioned in Markovian process, the
geometric phase acquired after a quasi-periods
$T=\frac{2\pi}{\omega}$ takes the same form as Eq.(\ref{GP}), where
$\theta_t$ and $\eta$ are defined by
\begin{eqnarray}
\cos^2\frac{\theta_t}2=\frac{(\sin^2\frac{\theta}2+\eta-\cos^2\frac{\theta}2e^{-2\gamma
t})^2}{(\sin^2\frac{\theta}2+\eta-\cos^2\frac{\theta}2e^{-2\gamma
t})^2+\sin^2\theta e^{-\gamma t}},
\end{eqnarray}
\begin{eqnarray}
\eta=\sqrt{\left((1+e^{-2\gamma
t})\cos^2\frac{\theta}2-1\right)^2+\sin^2\theta e^{-\gamma t}}.
\end{eqnarray}

\begin{figure}
\includegraphics*[width=0.9\columnwidth,
height=0.65\columnwidth]{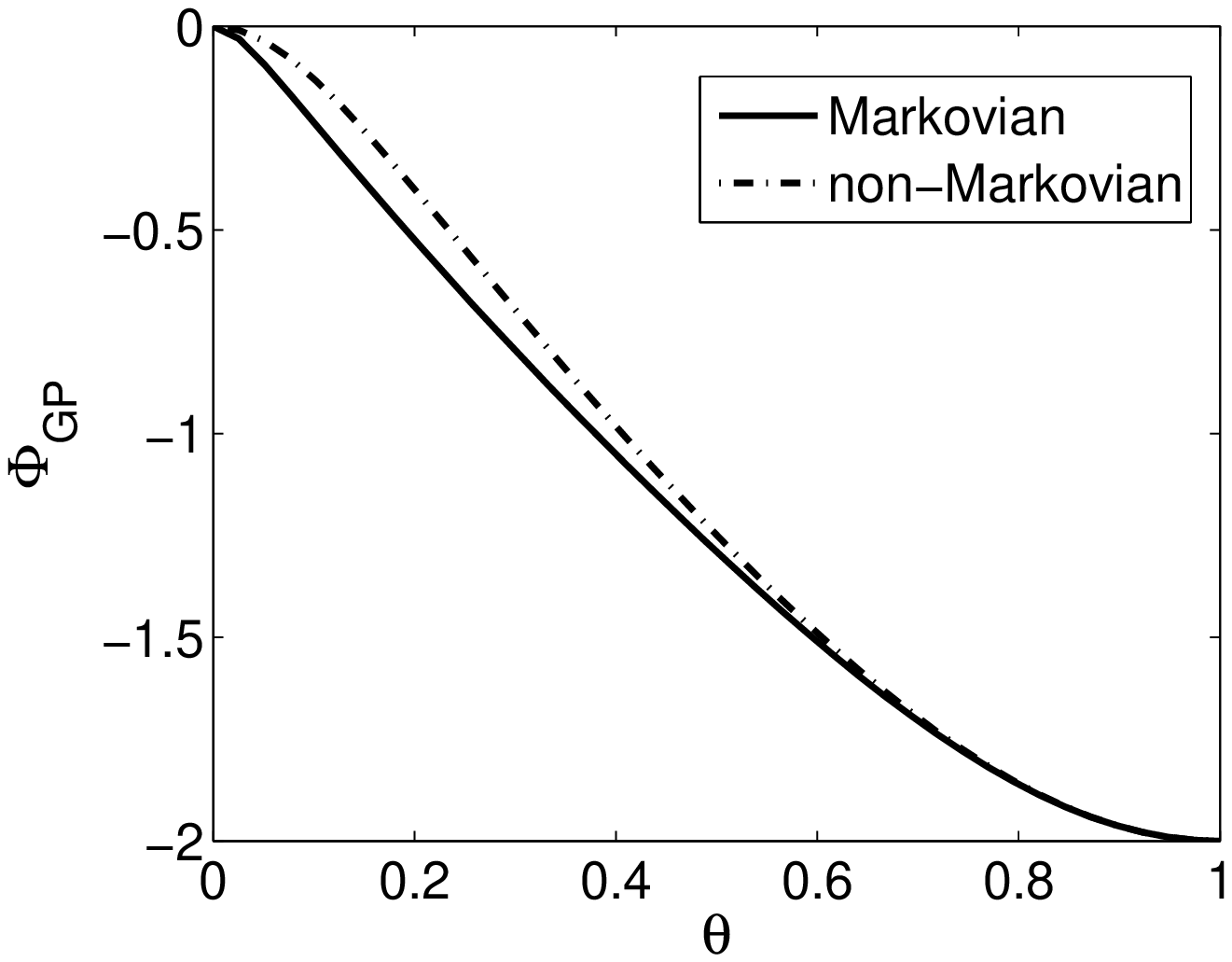}
\includegraphics*[width=0.9\columnwidth,
height=0.65\columnwidth]{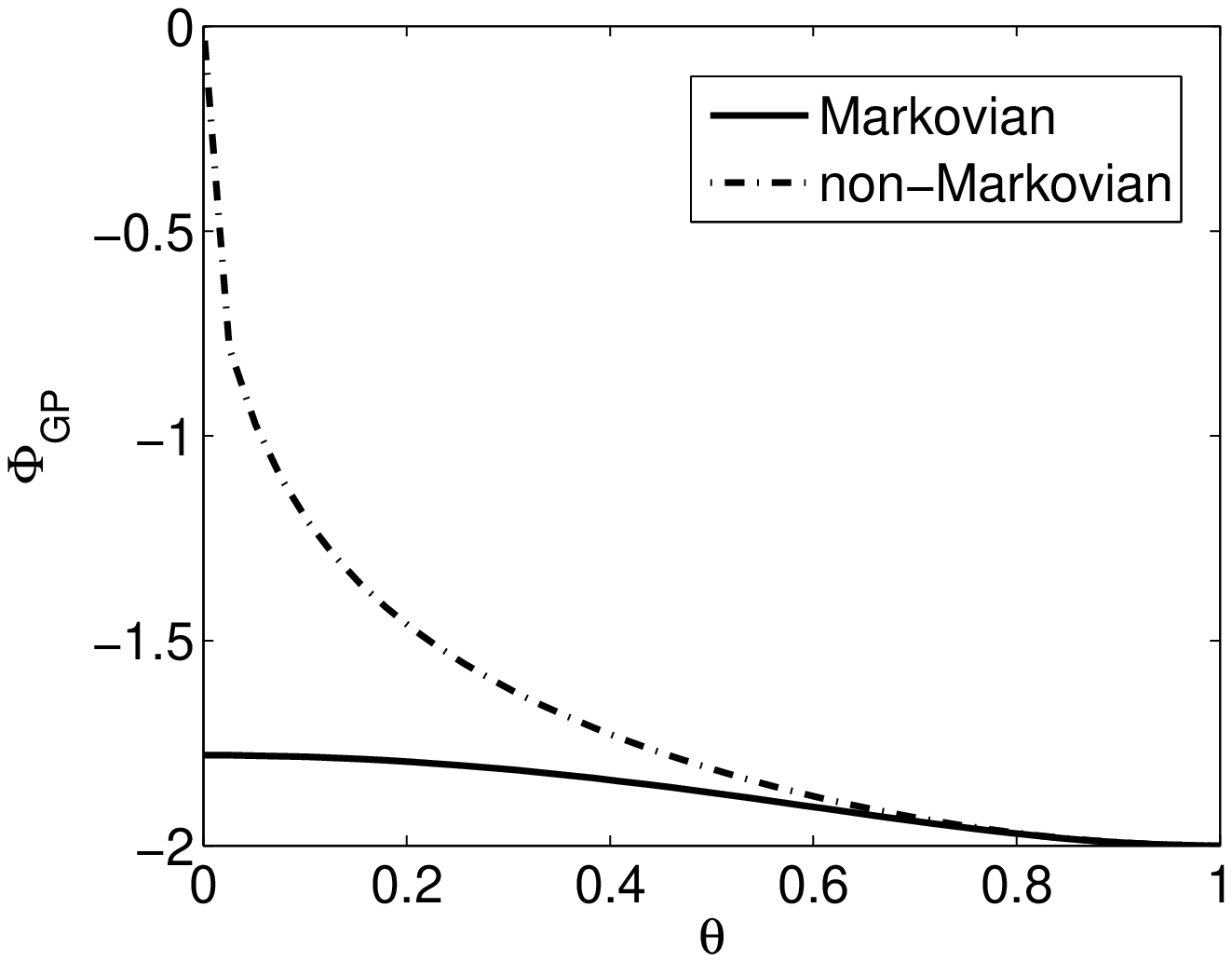} \caption{Geometric
phase in a system governed by the generalized Lindblad master
equation. The geometric phase $\Phi_{GP}$ and the polar angle
$\theta$ are plotted in units of $\pi$. We set the transition
frequency $\omega=1$ and time $\tau=\frac{2\pi}{\omega}$. The
parameter $\gamma$ is chosen as: top: $\gamma=0.1$, bottom:
$\gamma=1$. The solid line indicates the Markovian process while the
dash-dot line denotes the non-Markovian one.} \label{geophgLindblad}
\end{figure}

The geometric phases as a function of initial polar angle $\theta$
in both Markovian and non-Markovian cases are shown in
Fig.\ref{geophgLindblad}. In this figure, we have set transition
frequency $\omega=1$ and $\gamma_1=\gamma_2=\gamma$. The geometric
phase $\Phi_{GP}$ and the polar angle $\theta$ are plotted in units
of $\pi$. We see from this figure that when the dissipation is
weak($\gamma=0.1$), the geometric phase for both processes are
similar especially when $\theta>\frac{\pi}2$, whereas when the
dissipation is large($\gamma=1$), the non-Markovian effect changes
the geometric phase drastically, especially when
$\theta\rightarrow0$. Because the time dependent for the
off-diagonal elements $\rho_{12}$ and $\rho_{21}$ are the same, we
shall study the difference of diagonal elements $\rho_{11}$ in both
cases. The decay functions are $A_M=e^{-\gamma t}$ and
$A_N=\frac12(1+e^{-2\gamma t})$ for Markovian and non-Markovian
equation respectively. When $\gamma t$ is small enough, we can make
a Taylor expansion and ignore all items higher than the
second-order. Then $A_M=A_N=1-\gamma t$, i.e. the dynamics are the
same for weak dissipation and short time, so it can be easily
understood that the geometric phases are nearly the same. When the
dissipation is large, the items higher than the second-order can not
be ignored for a quasi-periods. In this case, we set
$\theta\rightarrow0$, i.e., the initial state vector in Bloch sphere
is near the $z$ axis. After a quasi-periods, the vector turns to the
hemisphere containing the $-z$ axis in Markovian case while it
remains in the initial hemisphere in the non-Markovian one. This is
the reason why the difference  is so large between both processes
with large dissipation rate at $\theta\rightarrow0$, with which the
geometric phase is usually interpreted as the solid angle of the
evolution track in Bloch sphere. We also note that in the figures
here and below in this paper, the left boundaries of $\theta$ are
all set $\theta\rightarrow0$. If $\theta=0$, the geometric phase in
all these processes are zero. The qualitative analysis (see the
solid line in the bottom figure) is following.In the Markovian case,
because of the semi-positivity of $\eta$, the two branches of the
eigenvalues $\lambda_{\pm}$ maintain the relation
$\lambda_+>\lambda_-$ all the time for $\theta\neq0$. The
calculation of the geometric phase given above is available. When
$\theta\rightarrow0$ and $\gamma t=\ln2$ the two eigenvalues tend to
be equal, i.e. the eigenstates of the density matrix are
approximately degenerate. After this critical point, the relation
$\lambda_+>\lambda_-$ holds again in the above calculation. However,
as a matter of fact, in the case of $\theta=0$, the two branches
should be crossing,  and the relation $\lambda_+>\lambda_-$ should
be changed after the critical point. Moreover, when the degeneration
occurs, the geometric phase should be calculated as discussed at the
end of Ref.\cite{TongDMGP}, which in our model shows that when
$\theta=0$, the matrix elements of $\rho(t)$ and the eigenvalues and
as well as the coefficients of the eigenstates are all real. Those
facts  make the geometric phase zero all the time. This results in
the difference in the two cases $\theta=0$ and $\theta\rightarrow0$.

\section{non-Markovian effects with exponential memory}

In this section, we will calculate the geometric phase of a
two-level atom by using a memory kernel master equation with
exponential memory phenomenologically. This equation may lead to a
non-positive  reduced density matrix\cite{expmemoryhazard} and the
positivity for a qubit has been discussed in
Ref.\cite{expmemorypositivity}. In this paper, we focus on the
geometric phase only and do not take the positivity into account.

We consider a two-level atom interacting with a vacuum. An
integrodifferential master equation containing the memory kernel in
the interaction picture is\cite{expmemoryhazard,expmemorypositivity}
\begin{eqnarray}
\frac{d\rho}{dt}=\int_0^tK(t-t')\mathcal
{L}\rho(t')dt'=\int_0^tK(t')\mathcal {L}\rho(t-t')dt',
\end{eqnarray}
where $\mathcal {L}$ is the Liouvillian superoperator that may take
the form
\begin{eqnarray}
\mathcal
{L}\rho=\frac12\gamma_0(2\sigma^-\rho\sigma^+-\sigma^+\sigma^-\rho-\rho\sigma^+\sigma^-),
\end{eqnarray}
and $K(t)$ is the memory kernel and here we choose an exponential
memory phenomenologically
\begin{eqnarray}
K(t)=\gamma e^{-\gamma t}.
\end{eqnarray}
We call $\tau_R=\frac1{\gamma}$ the memory time. $\gamma_0$ is the
dissipation constant. One can solve this integrodifferential
equation by taking its Laplace transform, determining the poles and
inverting the solution in the standard way. The analytic solution of
this equation\cite{qubitsolution} with initial condition
Eq.(\ref{initialstate}) in Schr\"{o}dinger representation  is
\begin{eqnarray*}
\rho(t)=\left(\matrix{\xi(R,\tau)\cos^2\frac{\theta}2 &
\frac12\sin\theta e^{-i(\phi+\omega t)}\xi(\frac{R}2,\tau) \cr
 \frac12\sin\theta e^{i(\phi+\omega t)}\xi(\frac{R}2,\tau) &
 1-\xi(R,\tau)\cos^2\frac{\theta}2\cr}\right),
\end{eqnarray*}
where $\xi(R,\tau)$ is given by Eq.(\ref{memeoryxi}),
\begin{widetext}
\begin{eqnarray}
\xi(R,t)=\exp\left(-\frac{\tau}2\right)\left\{\frac1{\sqrt{|1-4R|}}\sinh[\frac{\tau
}2\sqrt{|1-4R|}]+\cosh[\frac{\tau}2\sqrt{|1-4R|}]\right\}\label{memeoryxi},
\end{eqnarray}
\end{widetext}
and $R=\frac{\gamma_0}{\gamma}$, $\tau=\gamma t$. This coefficients
are valid for $4R<1$ and $2R<1$. For $4R>1$ and $2R>1$, the form of
these coefficients are acquired by substituting $\sinh[.]$ and
$\cosh[.]$ with $\sin[.]$ and $\cos[.]$.

After the same procedure given above, we obtain the geometric phase
acquired at time $T=\frac{2\pi}{\omega}$ as Eq.(\ref{GP}) with
\begin{eqnarray}
\cos^2\frac{\theta_t}2=\frac{(1+\eta-2\xi(R,\tau)\cos^2\frac{\theta}2)^2}{
(1+\eta-2\xi(R,\tau)\cos^2\frac{\theta}2)^2+\xi^2(\frac{R}2,\tau)\sin^2\theta},
\end{eqnarray}
and
\begin{eqnarray}
\eta=\sqrt{\left(1-2\cos^2\frac{\theta}2\xi(R,\tau)\right)^2+\xi^2(\frac{R}2,\tau)\sin^2\theta}.
\end{eqnarray}

\begin{figure}
\includegraphics*[width=0.9\columnwidth,
height=0.65\columnwidth]{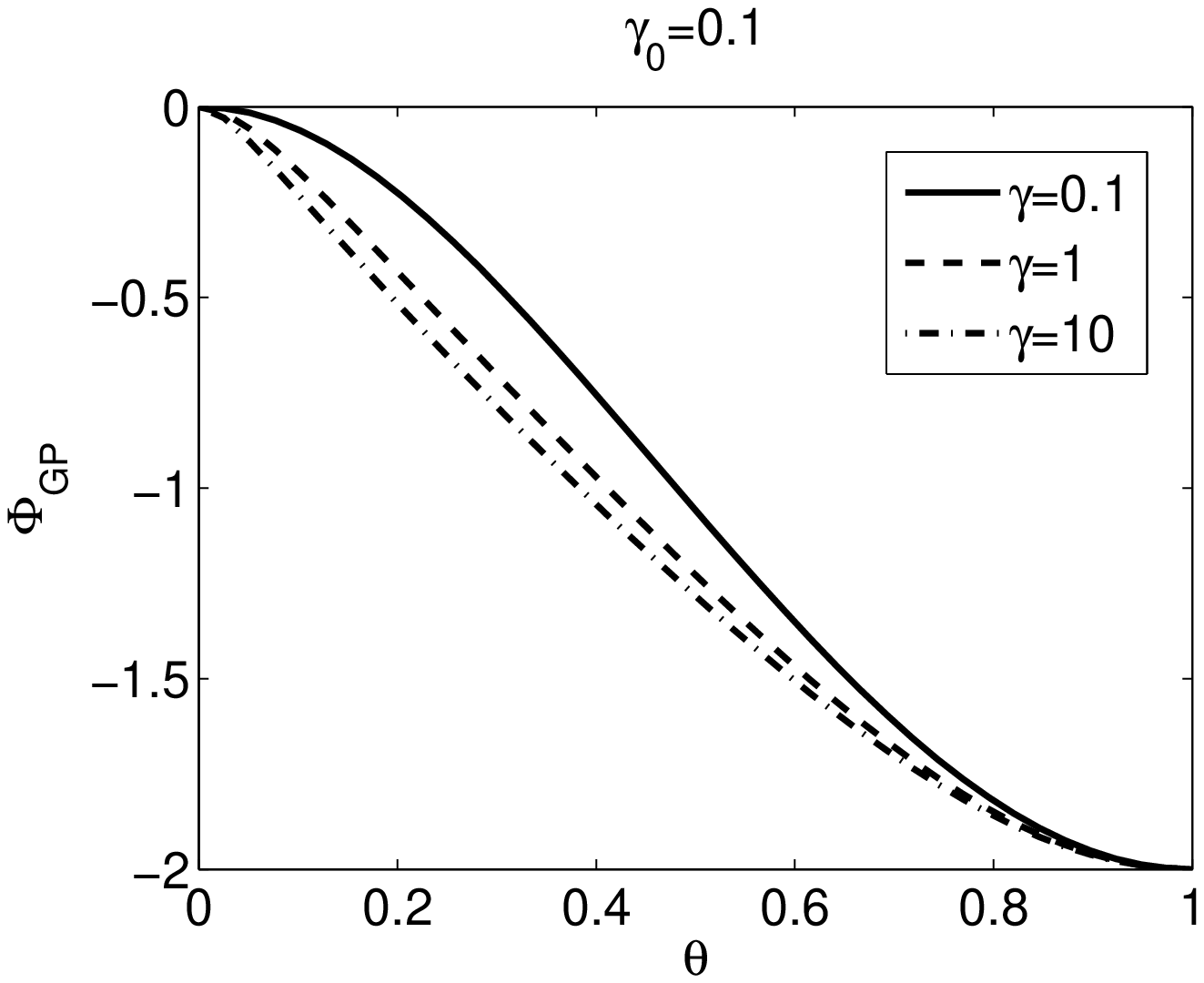}
\includegraphics*[width=0.9\columnwidth,
height=0.65\columnwidth]{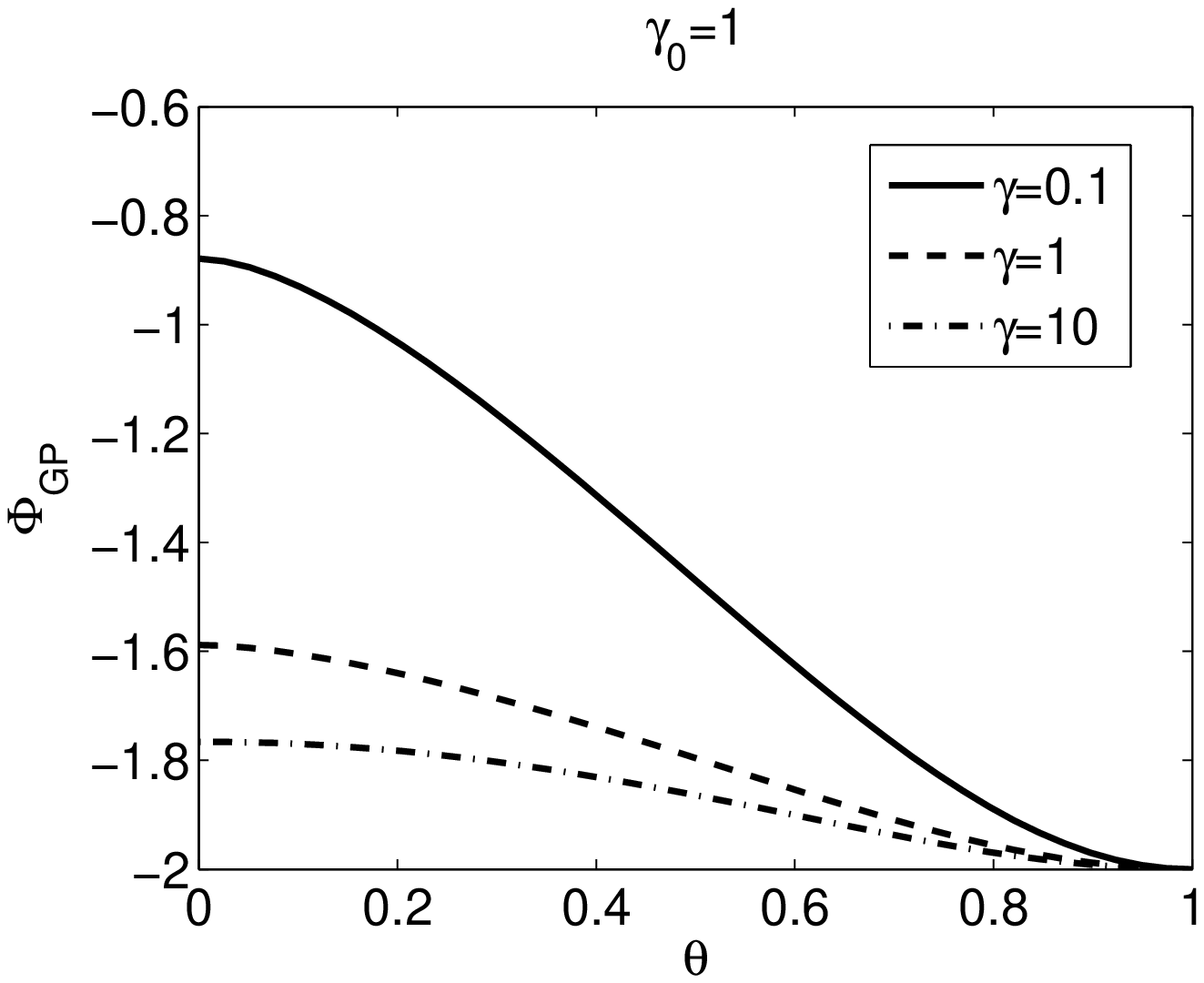} \caption{Geometric
phase of generalized Lindblad master equation with exponential
memory as a function of $\theta$. The geometric phase $\Phi_{GP}$
and the polar angle $\theta$ are plotted in units of $\pi$.
 The transition frequency $\omega=1$ and time
$\tau=\frac{2\pi}{\omega}$. The parameter $\gamma_0$ is chosen: top:
$\gamma_0=0.1$, bottom: $\gamma_0=1$. Solid line indicate
$\gamma=0.1$, dash line $\gamma=1$, and dash-dot line $\gamma=10$.}
\label{geophexpmemory}
\end{figure}

The geometric phases of the system governed by the generalized
Lindblad master equation as a function of $\theta$ are shown in
Fig.(\ref{geophexpmemory}), we plot this figure  with exponential
memory for different dissipation rate $\gamma_0$ and different
memory time $\gamma$. The geometric phase $\Phi_{GP}$ and the polar
angle $\theta$ are plotted in units of $\pi$. We see from the figure
that for weak dissipation(small dissipation
constant,$\gamma_0=0.1$), the three curves stand much close, even
when the memory time is long($\gamma=0.1$). For large dissipation
rate, long time memory lead to a large departure from the Markovian
process. When the memory time is shortened($\gamma=10$), the result
tends to the Markovian.

\section{Geometric phase with post-Markovian master equation}
In Ref.\cite{postMarkovian}, a post-Markovian master equation
including bath memory effects via a phenomenological introduced
memory kernel $K(t)$ is derived. Compared with the memory kernel
master equation, the advantage of post-Markovian master equation is
that for a qubit it keeps the positivity of the density matrix for
an exponential memory\cite{qubitsolution}. The general form of
post-Markovian master equation for a two-level atom in
zero-temperature reservoir is
\begin{equation}
\frac{d}{dt}\rho=\mathcal {L}\int_0^tK(t')e^{\mathcal
{L}t'}\rho(t-t')dt',
\end{equation}
where $\mathcal {L}$ is the Liouvillian superoperator and $K(t)$ is
the exponential memory kernel. It has been proved that the memory
kernel master equation is a special case of post-Markovian master
equation and one may derive the memory kernel master equation from
the post-Markovian equation in the limit
$\gamma_0\ll\gamma$\cite{qubitsolution}.

The time dependent of the state for atom and the geometric phase
after a quasi-periods have the same form as that from the memory
kernel master equation, the only difference is that the quantity
$\xi(R,\tau)$ is replaced by\cite{qubitsolution}
\begin{eqnarray}
\xi(R,\tau)=\frac{e^{-R\tau}-Re^{-\tau}}{1-R}.
\end{eqnarray}

\begin{figure}
\includegraphics*[width=0.9\columnwidth,
height=0.65\columnwidth]{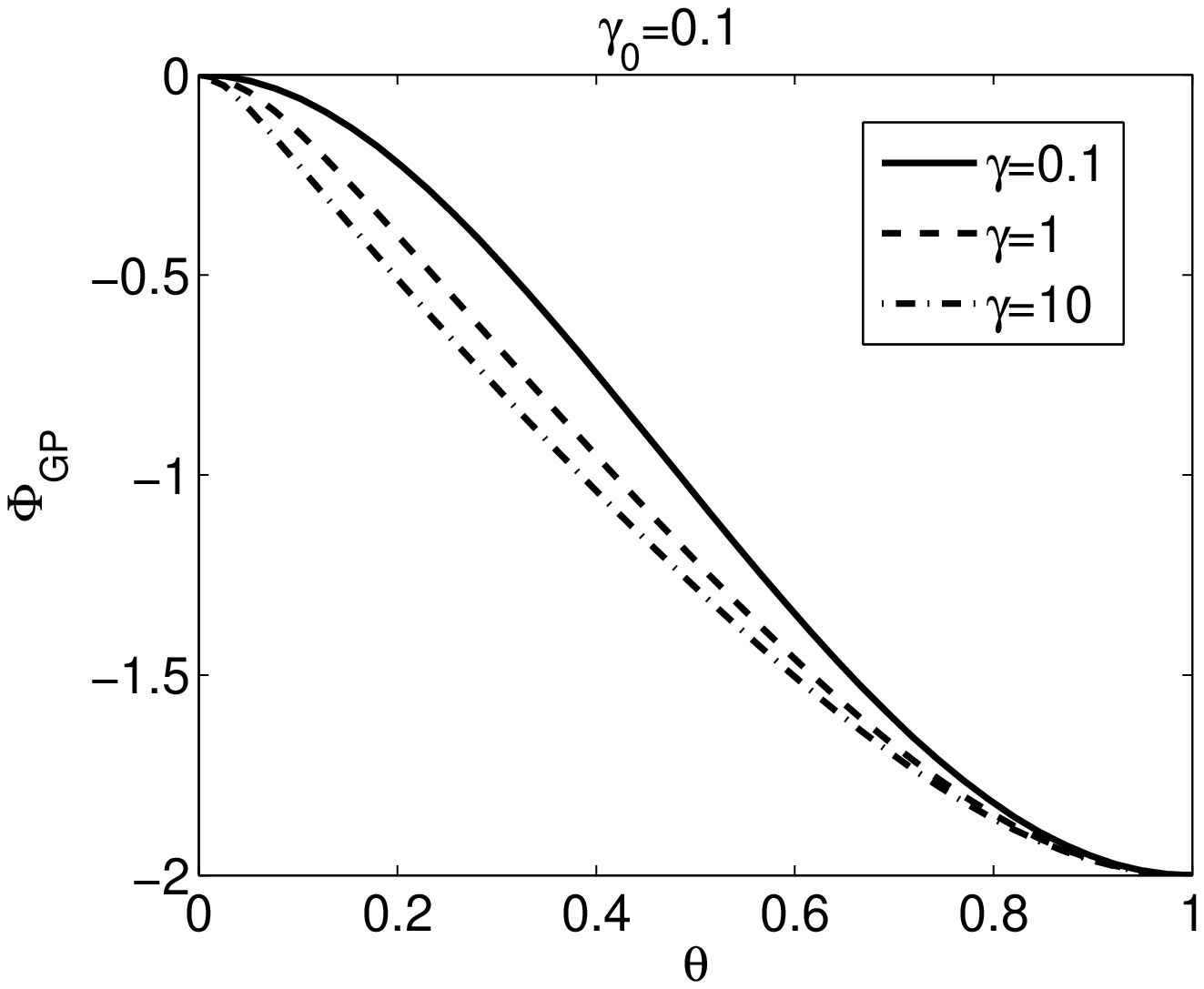}
\includegraphics*[width=0.9\columnwidth,
height=0.65\columnwidth]{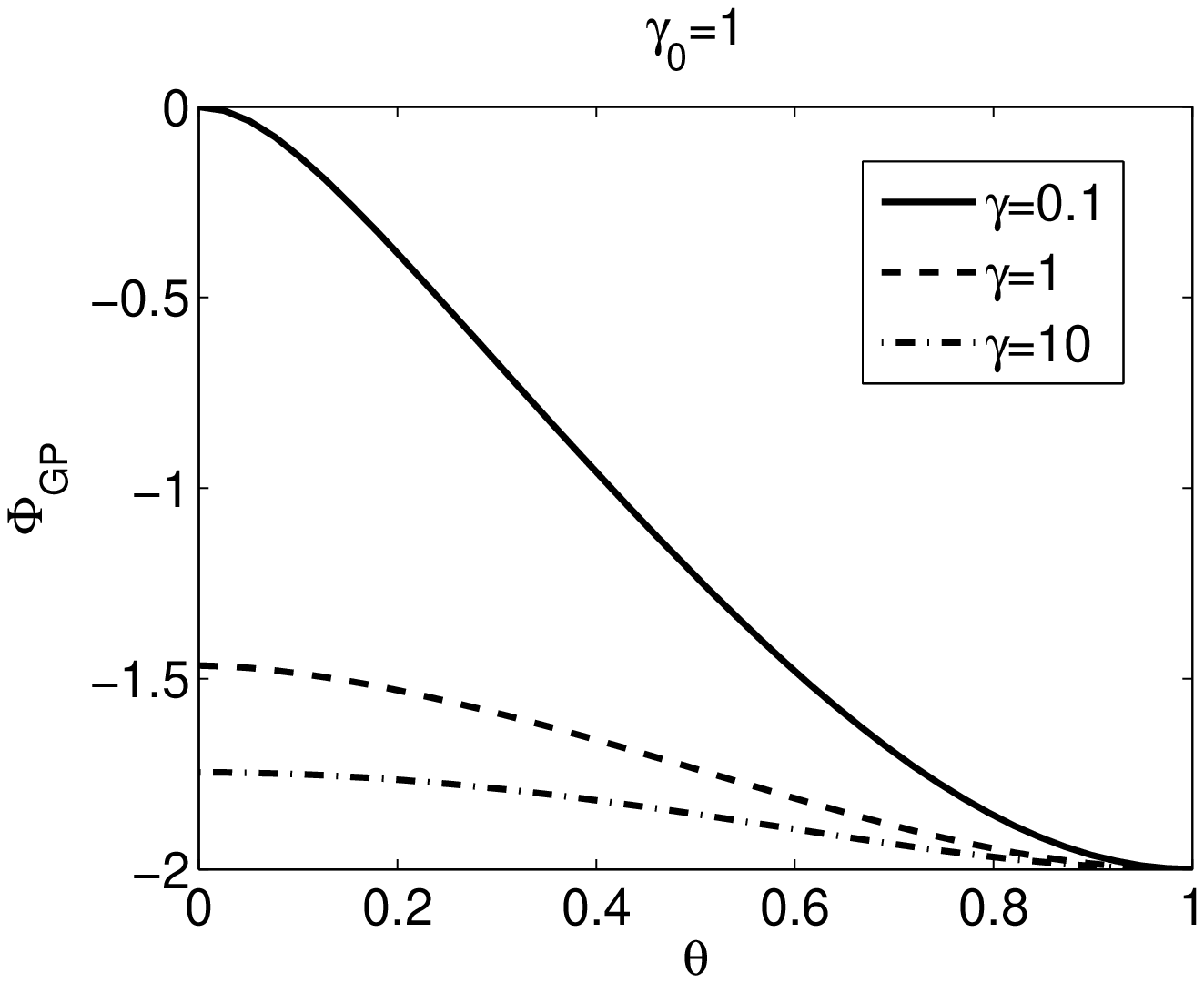} \caption{The same as
Fig.(\ref{geophexpmemory}) for post-Markovian master equation.}
\label{geophpost}
\end{figure}

The geometric phase for the post-Markovian master equation is shown
in Fig.(\ref{geophpost}). A similar feature  as that from the memory
kernel master equation can be seen in the figure. Furthermore when
the dissipation rate  is small, the figure is nearly the same as
that obtained from the memory kernel master equation. The difference
between the two cases can be found in the large dissipation case,
e.g., the geometric phase departure from the Markovian process more
largely with $\gamma=0.1$ than that in memory kernel master
equation.

\section{Conclusion}
In summary, we have studied the geometric phases of a two-level
system coupled to a non-Markovian environment. The non-Markovian
effects on the geometric phase are presented and discussed. We have
chosen three different methods to describe the non-Markovian
process. The common feature is that the geometric phase tends to
zero with $\theta$ approaches $\pi$. As the memory kernel master
equation can be reproduced  from the post-Markovian master equation
in the limit of $\gamma_0\ll\gamma$, the geometric phase calculated
confirmed this point in this limit. Our results also show that for
all the methods the non-Markovian effects change the geometric phase
more strikingly when the dissipation rate and population  of excited
state in the initial state is large. This can be understood as the
competition between the decoherence and the non-Markovian effects in
the open system.

\acknowledgments This work is supported
by NCET of M.O.E, and NSF of China under grant Nos. 60578014 and 10775023.\\

\end{document}